\documentclass[
    ,final            
epsfig  ]
  {aipproc}

\layoutstyle{6x9}

\begin{document}

\title{Studying the Sivers function by model calculations}

\author{A. Courtoy} {
address={
Departament de Fisica Te\`orica, Universitat de Val\`encia
and Institut de Fisica Corpuscular, Consejo Superior de Investigaciones
Cient\'{\i}ficas,
46100 Burjassot
(Val\`encia), Spain
}
}

\author{S. Scopetta}{
 address={
Dipartimento di Fisica, Universit\`a degli Studi
di Perugia,
via A. Pascoli,
06100 Perugia, Italy
}, 
altaddress=
{INFN, sezione di Perugia, via A. Pascoli,
06100 Perugia, Italy
}
}
\author{V. Vento}{
address= {
Departament de Fisica Te\`orica, Universitat de Val\`encia
and Institut de Fisica Corpuscular, Consejo Superior de Investigaciones
Cient\'{\i}ficas,
46100 Burjassot
(Val\`encia), Spain
},
altaddress={
TH-Division, PH Department, CERN, CH-1211 Gen\`eve 23, 
Switzerland
}
}

\begin{abstract}

A formalism is presented to evaluate the Sivers function
in constituent quark models.
A non-relativistic reduction of the
scheme is performed and applied to the Isgur-Karl model. 
The sign for the $u$ and $d$ flavor contributions that we obtained turns
out to be opposite. 
The Burkardt Sum Rule is fulfilled to a large extent.
After the estimate of the QCD evolution of
the results from the momentum scale of the model to the experimental
one, a reasonable agreement with the available data
is obtained.
\end{abstract}

\maketitle
The partonic structure of transversely polarized nucleons
is still an open problem\cite{bdr}.
Semi-inclusive deep inelastic scattering (SIDIS)
is one of the proposed
processes to access the parton distributions (PDs)
of transversely polarized hadrons.
SIDIS of unpolarized electrons off a transversely polarized target
shows "single spin asymmetries'' (SSAs) \cite{Collins},
due to two physical mechanisms,
whose contributions can be distinguished
\cite{mu-ta,ko-mu,boer}, i.e. the Collins\cite{Collins} 
and the Sivers\cite{sivers}  mechanisms.
The Sivers mechanism leads to a SSA which is the product of 
the unpolarized fragmentation function with  the Sivers PD. The latter
describes the number density of unpolarized quarks
in a transversely polarized target: it is a time-reversal odd,
Transverse Momentum Dependent  (TMD) PD.
From the existence of leading-twist
Final State Interactions (FSI) \cite{brohs,brodhoy}, 
 a non-vanishing Sivers function has been explained as 
 generated by the gauge link in the definition of TMDs
\cite{coll2,jiyu,adra}, whose
contribution does not vanish in the light-cone gauge,
as happens for the standard PD functions.
Different parameterizations of the available
SIDIS data have been published \cite{ans,coll3,Vogelsang:2005cs}, 
still with large error bars.
Since a calculation from first
principles in QCD is not yet possible, 
several model evaluations have been performed, e.g. in a quark-diquark model
\cite{brohs,bacch}; in the MIT bag model~\cite{yuan}; 
in a light-cone model \cite{luma}.
\begin{figure}
 \centering
  \includegraphics[height=.14\textheight]{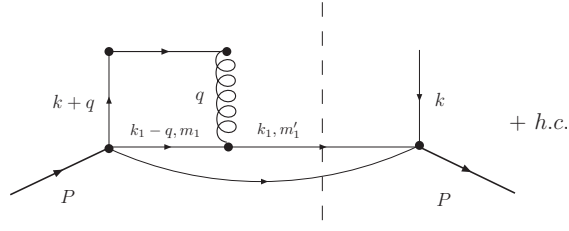}
  \caption{The contributions to the Sivers
function in the present approach.}
\end{figure}

We here describe a Constituent Quark Model (CQM) calculation 
of the Sivers function
\cite{nostro}.
CQM calculations of PDs are based on a two steps procedure\cite{trvv}.
First, the matrix element of the
proper operator is evaluated using the wave functions of the model;
then, a low momentum scale, $\mu_0^2$, is ascribed to
the model calculation and QCD evolution is used
to evolve the observable calculated in this low energy scale to
the  scale of DIS experiments.
Such
procedure has proven successful in describing the gross features of
PDs  
and GPDs\cite{h1},
 by using different CQMs, e.g. the Isgur-Karl (IK) model \cite{ik}. 
 Besides the fact that it successfully reproduces the low-energy properties of the nucleon, the IK model
 contains the one-gluon-exchange (OGE)
mechanism\cite{ruju}. 
In the present calculation,  the leading twist contribution to the
FSI has to be taken into account. We, here, consider the leading, OGE, order,
which is natural in the IK model. 
 The other approximations in our approach
are that, first, only the valence quark sector is investigated; 
second, that the resulting interaction is obtained
through a non-relativistic (NR) reduction of the relevant operator,
according to the philosophy of constituent quark models \cite{ruju}.
The Sivers function (Fig. 1) for  a proton polarized along the $y$ axis and for the
quark of flavor ${\cal Q}$ 
takes the form
\begin{eqnarray}
f_{1T}^{\perp {\cal Q}} (x, {k_T} )
& =&\Im
\left \{
- i g^2
{
M^2 \over k_x
}
\int
d \vec k_1
d \vec k_3
{d^2 \vec q_T \over (2 \pi)^2}
\delta(k_3^+ - xP^+)
\delta(\vec k_{3 T} + \vec q_T - \vec k_T) {\cal M}^{\cal Q}
\right \}\,\,
\label{start2}
\end{eqnarray}
where $g$ is the strong coupling constant, $M$ the proton mass,
and
\begin{eqnarray}
{\cal M}^{u(d)}  &=& 
\sum_{m_1,m_1',m_3,m_3'}
\Phi_{sf,S_z=1}^{\dagger}
\left ( \vec k_3, m_3; \vec k_1, m_1;
\, \vec P - \vec k_3 - \vec k_1,  m_n  \right )
\nonumber
\\
&&{ 1 \pm \tau_3(3) \over 2 }
V_{NR}(\vec k_1, \vec k_3, \vec q)
\nonumber
\\
&&
\,\Phi_{sf , S_z=-1}
\left (\vec k_3 + \vec q, m_3'; \, \vec k_1 -
\vec q, m_1';
\, \vec P - \vec k_3 - \vec k_1,  m_n  \right )~.
\label{Mu}
\end{eqnarray}
Using the spin-flavor wave function of the proton
in momentum space, 
$\Phi_{sf}$, corresponding
to a given CQM, the Sivers function,
Eq. (\ref{start2}),
can be evaluated.
From Eq.~(\ref{Mu}), one  notices that the helicity conserving
part of the global interaction
does not contribute to the Sivers function.
Besides,
in an extreme NR limit, it turns
out to be identically zero:
In our  scheme, it is precisely the interference of the lower
and upper components in the four-spinors
of the free quark states which leads to a non-vanishing
Sivers function. This holds even from
the component with $l = 0$ of the target wave function.
While, in other approaches\cite{yuan},  these interference terms arise due to the wave function,
they are produced here by the interaction.

The above-described formalism is now applied to the IK model.
\begin{figure}
\includegraphics[width=.4\textwidth]{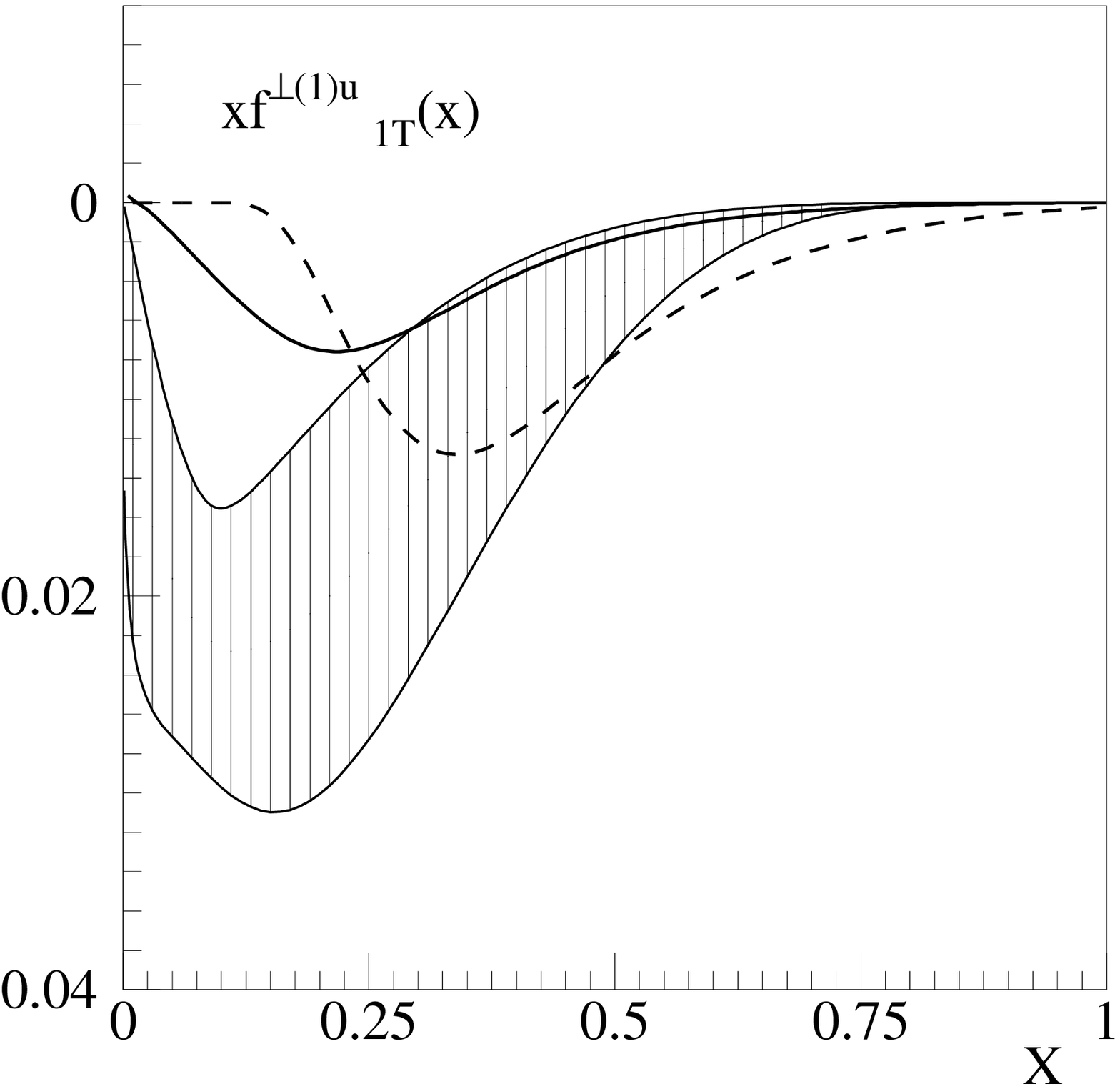}
\includegraphics[width=.4\textwidth]{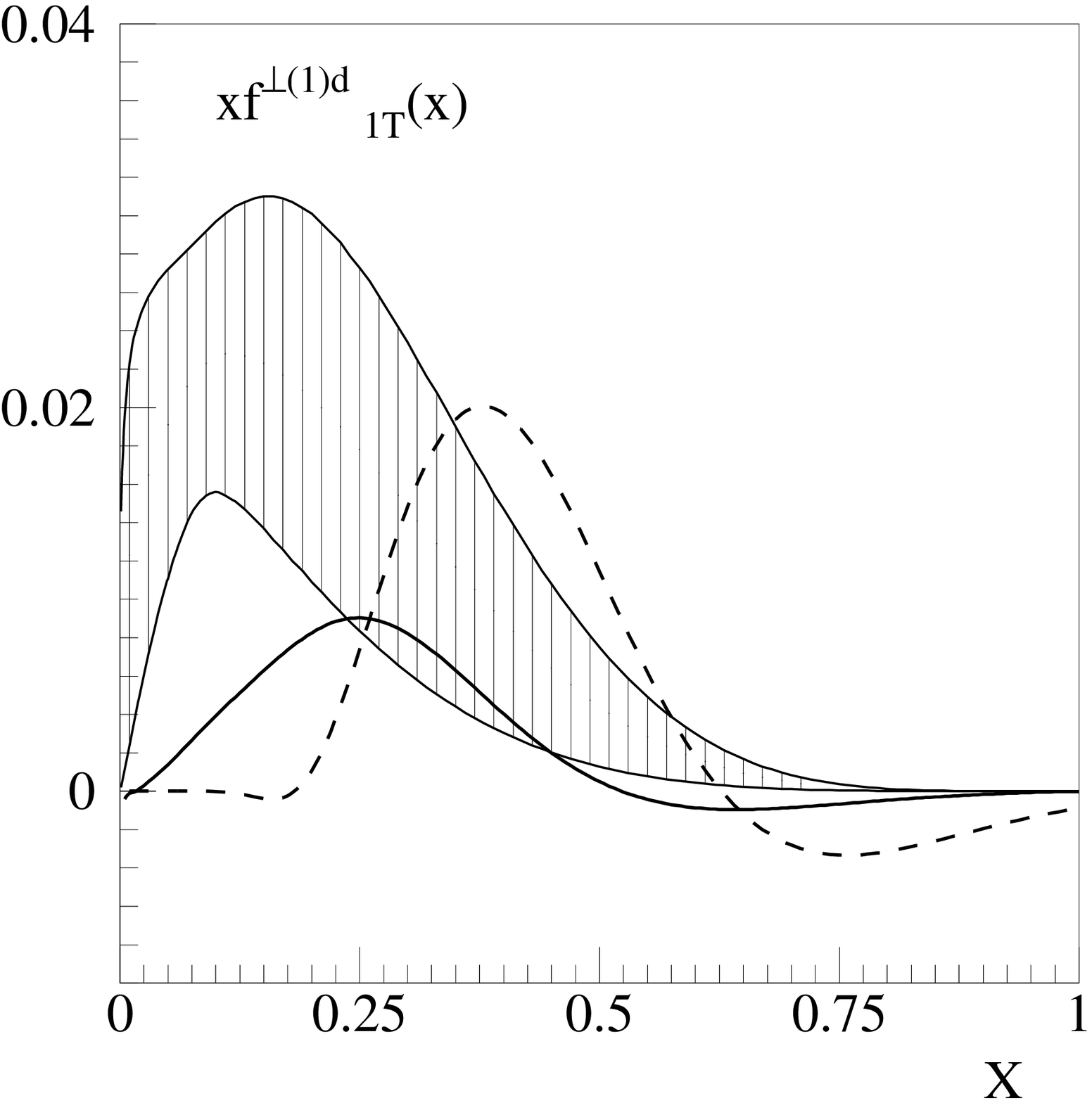}
\caption{
Left (right): the quantity $f_{1T}^{\perp (1) u(d) }(x) $, Eq. (\ref{momf}).
Dashed curve: IK at  $\mu_0^2$.
Full curve:  the evolved distribution at NLO.
Patterned area: parameterization by\cite{coll3} (see text).
}
\end{figure}
To evaluate numerically Eq. (\ref{start2}), $g$ (i.e.
$\alpha_s(Q^2)$) has to be fixed.
The prescription\cite{trvv} 
is used to fix $\mu_0^2$, according to the
amount of momentum carried by the valence quarks in the model.
Here, assuming that all the gluons and sea pairs in the proton
are produced perturbatively according to NLO evolution equations,
in order to have $\simeq 55 \% $ of the momentum
carried by the valence quarks at a scale of 0.34 GeV$^2$ 
one finds
that $\mu_0^2 \simeq 0.1$ GeV$^2$
if $\Lambda_{QCD}^{NLO} \simeq 0.24$ GeV.
This yields $\alpha_s(\mu_0^2)/(4 \pi) \simeq 0.13$ \cite{trvv}.
The results of the present approach
for the first  moments of the Sivers function, defined as
\begin{equation}
f_{1T}^{\perp (1) {\cal Q} } (x)
= \int {d^2 \vec k_T}  { k_T^2 \over 2 M^2}
f_{1T}^{\perp {\cal Q}} (x, {k_T} )~,
\label{momf}
\end{equation}
are given
by the dashed curves in Fig. 2.
They are compared
with a parameterization of the HERMES data,
taken at $Q^2=2.5$ GeV$^2$ :
The patterned area represents the $1-\sigma$ range
of the best fit proposed in Ref. \cite{coll3}.
The magnitude of the results
is close to that of the data,
although
they have a different shape: the maximum (minimum)
is predicted at larger values of $x$.
Actually $\mu_0^2$ is much lower, $Q^2 =2.5$ GeV$^2$. 
A proper comparison requires QCD evolution of TMDPDs,
 what is, to large extent, unknown.
We nevertheless perform a NLO evolution of the model
results assuming, for $f_{1T}^{\perp (1) {\cal Q} } (x)$,
the same anomalous dimensions of the unpolarized PDFs.
From the final result (full curve in Fig. 2),  one can see
that the  agreement with data
improves dramatically and the
trend is reasonably reproduced at least for $x \ge 0.2$.
Although the performed evolution is not exact, 
the procedure highlights the necessity 
of evolving the model results 
to the experiment scale and it suggests 
that the present results could be
consistent with data,  
still affected by large errors.

Properties of the Sivers function can be inferred from general principles. 
The Burkardt Sum Rule (BSR)
\cite{Burkardt:2004ur} states that, 
for a proton  polarized in the positive $y$ direction,
$\sum_{{\cal Q}=u,d} \langle k_x^{\cal{Q}} \rangle = 0$
with
\begin{equation}
\langle k_x^{\cal{Q}} \rangle = - \int_0^1 d x \int d \vec k_T
{k_x^2 \over M}  f_{1T}^{\perp \cal{Q}} (x, {k_T} )~,
\label{burs}
\end{equation}
and must be satisfied at any scale.
Within our scheme, at the scale of the model, it is found
$\langle k_x^{u} \rangle = 10.85$ MeV,
$\langle k_x^{d} \rangle = - 11.25$ MeV and, 
in order to have an estimate
of the quality of the agreement of our results with
the sum rule, we define the ratio
$r= 
| \langle k_x^{d} \rangle+
\langle k_x^{u}\rangle | /
| \langle k_x^{d} \rangle-
\langle k_x^{u} \rangle | $
obtaining $r \simeq 0.02$, so that we can say that our calculation
fulfills the BSR to a precision of a few percent.
One should notice that the agreement which is found
is better than that found in other model calculations
\cite{bacch,yuan},
especially for what concerns the fulfillment of the
Burkardt Sum Rule. However, in a recent work \cite{Courtoy:2008dn}, we have shown the encouraging result
 that the calculation in the
bag model satisfies the Burkardt sum rule at a 5$\%$ level.

\begin{theacknowledgments}
We thank A.~Prokudin for inviting us to present our results at this conference.
This work is supported by the INFN-CICYT agreement,
the Generalitat Valenciana (AINV06/118); the Contract No. 506078 (I3 Hadron Physics) and
 the MICINN (Spain) (FPA 2007-65748-C02-01, AP2005-5331 and PR2007-0048).
\end{theacknowledgments}

\bibliographystyle{aipprocl} 

\end{document}